\documentstyle[sprocl]{article}

\input epsf

\def\be{\begin{equation}}
\def\ee{\end{equation}}
\def\bea{\begin{eqnarray}}
\def\eea{\end{eqnarray}}

\def\slashchar#1{\setbox0=\hbox{$#1$}           
   \dimen0=\wd0                                 
   \setbox1=\hbox{/} \dimen1=\wd1               
   \ifdim\dimen0>\dimen1                        
      \rlap{\hbox to \dimen0{\hfil/\hfil}}      
      #1                                        
   \else                                        
      \rlap{\hbox to \dimen1{\hfil$#1$\hfil}}   
      /                                         
   \fi}                                         %

\def\simge{
    \mathrel{\rlap{\raise 0.511ex
        \hbox{$>$}}{\lower 0.511ex \hbox{$\sim$}}}}
\def\simle{
    \mathrel{\rlap{\raise 0.511ex 
        \hbox{$<$}}{\lower 0.511ex \hbox{$\sim$}}}}

\def\tg{{\tilde g}}

\def\tchi{{\tilde\chi}}

\def\lsp{{\tilde\chi_1^0}}
\def\tG{{\tilde G}}
\def\ttau{{\tilde\tau}}

\def\GeV{{\rm GeV}}

\def\sgn{\mathop{\rm sgn}}

\def\mhalf{m_{1/2}}

\def\dofig#1#2{\epsfxsize=#1\centerline{\epsfbox{#2}}}

\begin{document}

\font\twelvess=cmss10 scaled \magstep1

\begingroup
\parindent=20pt
\thispagestyle{empty}
\vbox to 0pt{
\vskip-1.4in
\moveleft0.75in\vbox to 8.9in{\hsize=6.5in

\centerline{\twelvess BROOKHAVEN NATIONAL LABORATORY}
\vskip6pt
\hrule
\vskip1pt
\hrule
\vskip4pt
\hbox to \hsize{August, 1999 \hfil BNL-HET-99/22}
\vskip3pt
\hrule
\vskip1pt
\hrule
\vskip3pt

\vskip1in
\centerline{\LARGE\bf SUSY EVENT GENERATORS}
\vskip.12in
\centerline{\LARGE\bf FOR LINEAR COLLIDERS}
\vskip.5in
\centerline{\bf Frank E. Paige}
\vskip4pt
\centerline{Physics Department}
\centerline{Brookhaven National Laboratory}
\centerline{Upton, NY 11973 USA}

\vskip.75in

\centerline{ABSTRACT}

\vskip8pt
\narrower\narrower
	The status of event generators for SUSY processes at future
Linear Colliders is briefly reviewed.

\vskip1in

	Invited talk at the {\sl International Workshop on Linear
Colliders (LCWS99)} (Sitges, Spain, 28 April -- 5 May 1999).

\vskip0pt

\vfil\footnotesize
	This manuscript has been authored under contract number
DE-AC02-98CH10886 with the U.S. Department of Energy.  Accordingly,
the U.S.  Government retains a non-exclusive, royalty-free license to
publish or reproduce the published form of this contribution, or allow
others to do so, for U.S. Government purposes.

\vskip0pt} 
\vss} 

\newpage
\thispagestyle{empty}
\
\newpage
\endgroup
\setcounter{page}{1}

\title{SUSY EVENT GENERATORS FOR LINEAR COLLIDERS}
\author{Frank E. Paige}
\address{Brookhaven National Laboratory, Upton, NY 11973, USA}
\maketitle

\abstracts{The status of event generators for SUSY processes at future
Linear Colliders is briefly reviewed.}

\section{Introduction}

	Parton shower Monte Carlo programs, reviewed by Sj\"ostrand in
the previous talk,\cite{TS} are important tools for relating
perturbative QCD cross sections to experiment. They also make possible
more realistic predictions of the potential for future accelerators to
study new physics. SUSY at the TeV mass scale is a well motivated
extension of Standard Model. A Monte Carlo approach is needed not only
to incorporate properly the effects of QCD but also to handle
efficiently the many possible chains of decays. Rather detailed
treatments of SUSY are now included in three parton shower Monte Carlo
programs:
\begin{itemize}
\item ISAJET 7.44,\cite{isajet} which incorporates ISASUSY by Baer and
Tata;
\item PYTHIA 6.1,\cite{pythia} which incorporates SPYTHIA by Mrenna;
\item HERWIG 6.1,\cite{herwig} which takes part of its treatment of
SUSY from ISAJET but adds especially $R$-parity violation.
\end{itemize}
\noindent This talk describes the physics assumptions made in these
event generators. It also discusses a few recent developments, such as
the incorporation of matrix elements in ISAJET.

\section{Minimal Supersymmetric Standard Model}

	The starting point for all three SUSY generators is the
Minimal Supersymmetric Standard Model (MSSM), that is, the SUSY model
with the same gauge group as the Standard Model and the minimal
particle content. Each chiral $f_{L,R}$ has a scalar partner $\tilde
f_{L,R}$, and each gauge boson has a spin-1/2 gaugino partner. There
are also two Higgs doublets with spin-1/2 Higgsino partners. 

	The most general such model allows weak scale proton decay.
The simplest solution is to impose symmetry under $R$-parity, where
$R=(-1)^{3(B-L)+2S}$. If $R$ is conserved, then SUSY particles
produced in pairs and the lightest SUSY particle (LSP) is stable. It
is also possible to avoid proton decay by assuming only $B$ or $L$
conservation. The $R$-violating couplings are Higgs-like, and some but
not all of them must be small.\cite{dreiner} If they are all small,
then the only effect of $R$ violation is to allow the LSP to decay,
e.g., $\lsp \to qqq$, $\lsp \to \ell^+\ell^-\nu$, or $\lsp \to q\bar
q\ell$. Larger couplings would also affect other SUSY decays; so far
this has been incorporated only in HERWIG.

	SUSY must of course be broken. It is possible to add
gauge-invariant masses for all the SUSY particles by hand, but at the
cost of 105 (or more) new parameters. Presumably these masses should
be generated by some sort of spontaneous symmetry breaking. This seems
to be impossible using only the MSSM, so one must introduce a hidden
sector to break SUSY at a mass-squared scale $F$ and communicate that
breaking to the MSSM at a scale $M$. In supergravity (SUGRA) models,
$M \sim M_{\rm Planck} = (8\pi G_N)^{-1/2} =2.4 \times 10^{18}\,\GeV$
so $F \sim 10^{11}\,\GeV$. In Gauge Mediated SUSY Breaking (GMSB)
models, the communication is via Standard Model gauge interactions,
and $M$ and $F$ are much smaller. This can give a light
($M_\tG \ll 1\,\GeV$) gravitino into which all other SUSY particles
decay.

	Electroweak symmetry is broken through the Higgs mechanism,
leaving five physical Higgs bosons, $h, H, A, H^\pm$. The gauginos and
Higgsinos then mix to give four neutralinos $\tchi_i^0$ and two
charginos $\tchi_i^\pm$. In most cases, the $\lsp$, $\tchi_2^0$,
$\tchi_1^\pm$ are mainly gauginos, while the heavier states have
Higgs-like couplings. The scalar fermions also mix, especially for the
third generation.

	A random choice of the 105 MSSM parameters will produce flavor
changing neutral currents and $CP$ violation. These can be avoided by
assuming that all phases and flavor mixings are absent. This produces
what is called in ISAJET the MSSM parameter set:  $M_{\tg}$, $\mu$,
$M_A$, $\tan\beta$, $M_{Q_1}$, $M_{d_R}$, $M_{u_R}$, $M_{L_1}$,
$M_{e_R}$, $M_{Q_3}$, $M_{b_R}$, $M_{t_R}$, $M_{L_3}$, $M_{\tau_R}$,
$A_t$, $A_b$, $A_\tau$, $M_{Q_2}$, $M_{s_R}$, $M_{c_R}$, $M_{L_2}$,
$M_{\mu_R}$ $M_1$, and $M_2$. This is a plausible starting point but
has no good theoretical justification.

	More tractable parameter sets can be obtained by assuming some
sort of universality. If all scalar and all gaugino masses are taken
to be degenerate at the GUT scale, then one obtains the minimal SUGRA
model. Electroweak symmetry breaking is driven by the large top Yukawa
coupling, leading to a four parameters:  $m_0$, $\mhalf$, $A_0$, and
$\tan\beta={v_u/v_d}$, plus $\sgn\mu=\pm1$. This model is incorporated
using a numerical solution to the renormalization group equations in
ISAJET and HERWIG and an analytic approximation in PYTHIA. Recently
optional additional parameters have been added to ISAJET so that one
can study deviations from it.

	An alternative is the minimal GMSB model, in which the
messenger sector contains $N_5$ $5+\bar5$ multiplets of $SU(5)$. There
are again just four parameters, the messenger scale $M$,
$\Lambda=F/M$, $N_5$, and $\tan\beta$. Gaugino masses are $\propto
\alpha \Lambda N_5$, while scalar squared masses are $\propto \alpha^2
\Lambda^2 N_5$, where $\Lambda=F/M$. The gravitino mass, or rather the
lifetime for gravitino decays, is also a parameter. ISAJET also
incorporates a few additional GMSB parameters.

\begin{figure}[t]
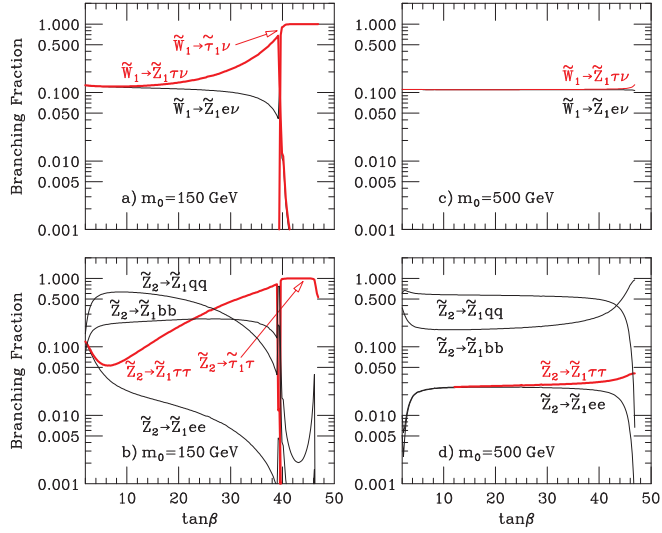

\dofig{3.4in}{nfig02.ai}
\caption{Gaugino branching ratios vs.\ $\tan\beta$ in the minimal SUGRA
model. \label{taus}}
\end{figure}

\section{\boldmath SUGRA with $\tan\beta \gg 1$}

	If $\tan\beta$ is much larger than one, then effects such as
$\tilde b_L-\tilde b_R$ and $\tilde\tau_L-\tilde\tau_R$ mixing become
important. This has been properly included in ISAJET since version
7.32. The most important effect is to increase the splitting of the
third generation masses. In particular, the $\ttau_1$ can become
light, so that the only two-body gaugino decay modes may be $\tchi_2^0
\to \ttau_1^\pm\tau^\mp$ and $\tchi_1^\pm \to \ttau^\pm\nu_\tau$. These
branching ratios may therefore be greater than 99\%, as can be seen in
Figure~\ref{taus}.\cite{bcdpt}

\begin{figure}
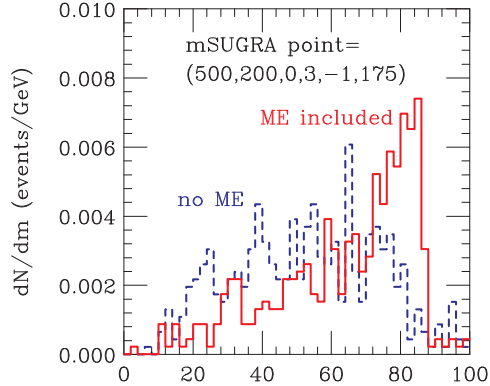

\dofig{2.5in}{me1.ai}
\caption{Dilepton mass distribution with (solid) and without (dotted)
matrix element. \label{me1}}
\end{figure}

	A limited amount of work has been done on such cases for the
LHC.\cite{me} While measurements are possible, they are clearly quite
difficult. Measurements at a linear collider may be easier, at least
relatively. E.g., one might measure masses using a threshold scan rather
than from $\tau$'s hadronic decays. This deserves more study.

\section{Matrix Elements in SUSY Decays}

	Three-body decays are often important for SUSY. While matrix
elements have always been included in the calculation of branching
ratios, they have mostly been ignored in the generation of events.

	The newest version of ISAJET incorporates matrix elements for
decays of the form $\widetilde A \to \widetilde B f \bar f$. One does
not want to duplicate all the code used to calculate branching ratios
inside the event loop. However, the most general matrix element has
only spin-0 poles in the $\widetilde B f$ and $\widetilde B \bar f$
channels and spin-0 and spin-1 poles in the $f \bar f$ channel. The
strategy is to save the relevant masses and couplings when calculating
the branching ratios and then to reconstruct the matrix element with a
single generic routine. The same strategy should work for HERWIG and
PYTHIA.

	The effect can be quite significant, as seen from the dilepton
mass distribution from $\tchi_1^\pm \tchi_2^0 \to \ell\nu\lsp
\ell\ell\lsp$ at the Tevatron shown in Figure~\ref{me1}. Nogiri and
Yamada\cite{nogiri} have also emphasised the importance of
including matrix elements. Incorporating spin correlations in a
general way is more difficult and has not been attempted. 

\section{ISAJET Bremsstrahlung/Beamsstrahlung}

\begin{figure}[t]
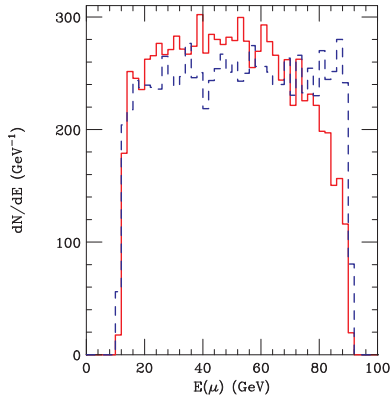

\dofig{2in}{emuon.ai}
\caption{Muon energy distribution with (solid) and without (dashed)
bremsstrahlung and beamsstrahlung. \label{emuon}}
\end{figure}

	Bremsstrahlung has been included in ISAJET~7.44 using the
Fadin-Kuraev $e^-$ distribution function. Beamsstrahlung has also been
implemented using the formalism of Chen, et al.\cite{cbp} The user
must specify $\Upsilon$ and $\sigma_z$. Both have significant 
effects on visible endpoint distributions, e.g., for
$e^+e^- \to \tilde\mu^+\tilde\mu^-$, as can be seen from
Figure~\ref{emuon}.

\medskip

	This work is supported in part by the U.S. Department of Energy
under contract DE-AC02-98CH10886.

\end{document}